\documentclass[a4paper,11pt]{article}
\usepackage{cite}
\usepackage{algorithm}
\usepackage{algorithmicx}
\usepackage{graphicx}
\usepackage{mathtools}
\usepackage[noend]{algpseudocode}
\usepackage{adjustbox}
\usepackage{url}
\usepackage{multirow}
\usepackage{subcaption}
\usepackage{fullpage}
\usepackage{authblk}

\newtheorem{problem}{Problem}
\newtheorem{definition}{Definition}

\begin{document}
%
\title{LP WAN Gateway Location Selection Using Modified K-Dominating Set Algorithm\thanks{This work has been partially funded by the Polish National Center for Research and Development grant no. POIR.04.01.04-00-0005/17.}}
%
%
\author[a]{Artur Frankiewicz}
\author[b]{Adam Glos}
\author[b]{Krzysztof Grochla} 
\author[a]{Zbigniew {{\L}}askarzewski}
\author[b]{Jaros{{\l}}aw Miszczak}
\author[b]{Konrad Po{{\l}}ys}
\author[b]{Przemys{{\l}}aw Sadowski}
\author[b]{Anna Strzoda}

\affil[a]{AIUT Sp. z o.o.,
Wycz{{\' o}}{{\l}}kowskiego 113, Gliwice, Poland}

\affil[b]{Institute of Theoretical and Applied Informatics, Polish Academy of Sciences, Ba{{\l}}tycka~5, Gliwice, Poland}

\date{October 9, 2020}

\maketitle              

\begin{abstract}
The LP WAN networks use gateways or base stations to communicate with devices distributed on large distances, up to tens of kilometres. The selection of optimal gateway locations in wireless networks should allow providing the complete coverage for a given set of nodes, taking into account the limitations, such as the number of nodes served per access point or required redundancy. In this paper, we describe the problem of selecting the base stations in a network using the concept of $k$-dominating set. In our model, we include information about the required redundancy and spectral efficiency. We consider the additional requirements on the resulting connections and provide the greedy algorithm for solving the problem. The algorithm is evaluated in randomly generated network topologies and using the coordinates of sample real smart metering networks.
\vspace{6pt}

\end{abstract}
\section{Introduction}
The introduction of Low Power WAN network concept, with devices communicating over distances of tens of kilometers using low power radio interface, has generated multitude of novel use cases based on long life battery powered devices. The LoRa (Long Range) is one of the most widely adopted LP WAN standards, which is based on ISM (Industrial, Scientific and Medical) frequencies and chirp spectrum modulation. The LoRa communication now is being deployed in devices such as smart meters, sensors or actuators executing smart city functions. The LoRa can provide cheap and very energy efficient communication with thousands of devices per one access point, at the cost of low data rate. To facilitate the deployment of LoRa devices, the LoRaWAN standard has been proposed, which defines the packet format and communication architecture allowing to forward the information from the devices to the internet servers. 

Traditionally radio planning technologies are used to select location of base stations in long range wireless networks. This approach is not always feasible in LP WAN, as due to the low cost of access point (few hundred dollars) and the simplicity of its deployment it is more efficient to deploy the access points in less efficient manner, but without the costly radio planning procedures. Traditionally, the gateway location selection problem was solved by radio planning engineers, using radio signal propagation models and simulation software and tested through the drive tests\cite{laiho2002radio}, \cite{hurley2002planning}. This method is to expensive for the LP WAN networks, which employ very large number of low cost devices, as it requires both engineering time to plan the radio and execution of time consuming drive tests. The number and locations of gateways need to be dynamically modified basing on the growth of the network, to provide coverage appropriate to the endpoints' density, as more end nodes generate more traffic which needs to be served. 

In this work, we propose an algorithm to select the suboptimal location for the LP WAN gateway location for a given set of endpoint location. The gateway location is selected from a set of location candidates, which may be the same as the set of end node location. The selection of a set of gateways must meet the requirement to provide connectivity to all end nodes and to provide the required redundancy. We assume that the connectivity graph estimated from the distances between the nodes is the input. We treat the problem as a problem of selection of a dominating set, which is an NP-complete decision problem in computational complexity theory \cite{alber2004polynomial}. Thus the problem is complex and it is hard to solve.

There are a few differences between other popular radio networks (e.g. LTE) and LoRa. LoRa end devices are typically battery powered devices so they transmit data rarely, but usually in a constant time window. LoRa uses ALOHA media access control so collisions are more probable. LoRa devices also use all possible channels uniformly, there is no "colouring" of areas and LoRa uses spreading factors, thus the capacity of the gateway may vary depending on the spatial distribution of the nodes. We assume the set of client locations and the traffic characteristics are known and is an input for the proposed algorithm, which is a case e.g. for advanced metering infrastructure (AMI) networks.

The rest of the paper is organized as follows: the second section describes the state of the art regarding the planning and dimensioning of LP WAN networks and the selection of AP location. Next, we describe the network model used in the study. In the following section, we show the pseudocode of the proposed optimization algorithm. Next we present the analysis of the performance of the algorithm for different types of topologies. We finish the paper with a short conclusion. 

\section{Literature review}

The radio network planning has been investigated for more than twenty years, however most of the work is dedicated to cellular networks - see e.g. \cite{laiho2002radio} or \cite{hurley2002planning}. However, there is very little research results available considering the specifics of the LP WAN radio network planning, where the cost of network deployment is much smaller, thus it does not justify the costly drive tests and radio signal propagation analysis. One of the few tools which allow to estimate the coverage of a LoRa gateway on a map is CloudRF \cite{cloudrf}. There are some other tools allowing to simulate radio signal propagation, of which most support ISM frequencies and allow to estimate the range of LP WAN access points, such as e.g. Radio Mobile Online \cite{rmo}. These tools are very useful in analysis of network coverage for a given set of access points, but require large knowledge of RF signal propagation and do not select automatically the gateway location, requiring the experience of the operator. While this is effective for a small number of access points, for large networks which need to match the restraints of number of nodes services by an access point use of such tools is ineffective. In the previous work by part of the authors \cite{grochla2020HeuristicAlgorithm} two of us have proposed a heuristic algorithm for gateway location selection, which however do not support redundancy.

\section{Gateway location selection procedure}

\subsection{Problem formulation}

To describe the problem of selecting the set of gateways for the given set 
of stations (end points), we assume that we have for our disposal the graph describing the 
visibility between the stations. In such graph two stations are connected only 
if it 
is possible to transmit packets between them. We assume that different 
connections may have different costs. We also 
make some assumption on capabilities of the resulting network. First, we
assume that each gateway has a fixed maximal capacity, the same for each 
of them, representing the maximal number of stations it can connect to 
simultaneously. This ensures that the transmission load will not exceed the 
capacity of the gateway. Second, we assume that each non-station has to be 
connected at least to some fixed number of gateways. This condition 
describes the redundancy requirements for the resulting network. Finally, we 
would like to find the smallest possible set of gateway, which expresses 
the requirement of using the smallest possible number of gateways.

To represent this situation we use a triple $(G,c,k)$, where $G$ is undirected 
weighted graph $G=(V,E)$, with the set of vertices $V$ and the set of edges 
$E$, $c$ is the maximal capacity of the gateways, and $k$ is the 
domination number describing the minimal number of gateways each 
station is connected to. In this description each vertex represents a station, 
and stations are connected if they are 
within their radius of visibility. To each edge, 
$(v,w)\in E$, we assign a weight $C_{vw}$, representing the cost of the 
connection between the station represented by vertices $v$ and $w$. The connection cost between nodes $v$ and $w$ is equal to $\displaystyle{\frac{1}{2^{12-SF_{vw}}}}$, where $SF_{vw}$ is equal to the value of spreading factor in the communication between nodes $v$ and $w$. Values of SF in the communication between two nodes are determined on the basis of a signal propagation model that takes into account the distance between nodes and random attenuation.
The higher the SF value, the higher the transmission cost. 
The domination number $k$ is used for representing the required 
redundancy.

\subsection{Description of the solution}

To describe our approach we first consider a simplified sub-problem, 
defined as follows.
\begin{problem}[Gateways]
	For a given set of stations, described by $(G,c,k)$, find the set $D$ of 
	stations such that each station is in this set or it has a direct 
	connection to a station from this set.
\end{problem}

One can note that, as the only requirement here is to assure the visibility 
between the stations, the solution of the above problem is given a subset of 
stations. Thus, no information about the connections is given.

The stations from the set $D$, obtained as a solution of the above problem, are 
called gateways. The set $D$ is called a dominating set of $G$. 
If we impose the condition that each element station 
should be connected to at least $k$ gateways, the solution of the above 
problem is equivalent to finding $k$-dominating set, $D_k$, of a graph $G$.

\begin{definition}
	For a graph $G=(V,E)$, the $k$-dominating set of $G$, denoted by $D_k$, is 
	any subset of vertices such that each vertex from $V$ is connected to at 
	least $k$ nodes from $D_k$ or belongs to~$D_k$.
\end{definition}

The notion of $k$-dominating set captures the redundancy requirement.  Still 
the information about the $D_k$ alone is not sufficient to describe the 
allocation of connection between stations and gateways.

A similar approach has been used previously for constructing routing algorithms 
in \emph{ad hoc} networks \cite{das1997routing}, and various methods have been 
developed in this context \cite{hochbaum1985best, wan2004distributed, 
	dai2005constructing}. However, one should keep in mind that in our scenario 
we assume that not only the resulting network has to be redundant, but also 
the resulting gateways have limited capacities. On the other hand, in 
contrast to the routing problem, we do not require the subgraph with vertex 
set consisting of gateways to be connected.

One should note, that if we assume the capacity $c$ in our problem to be 
infinite, then an arbitrary connection scheme with a gateway $D_k$ would 
be a proper solution. In particular, gateway can serve all of the neighbouring nodes. However, since capacity is finite, each gateway 
cannot serve all neighbouring nodes, thus we may need to add gateways in order to serve not-served one. Thus, the solution of our problem is given 
by the solution of $k$-domination problem, which is to find the minimal $k$-dominating set, with fixed capacity $c$.

Taking all this into account, we define our problem as follows.

\begin{problem}[Gateways with constrains]
	For a given graph, $G=(V,E)$, find the set of gateways $D_k$, forming a 
	$k$-dominating set $D_k\subset V$, and a graph, $H=(V,E_H)$, which is a 
	subgraph of $G$, $H\subset G$, satisfying the following conditions,
	\begin{enumerate}
		\item each edge from $E_H$ connects gatewas and stations,
		\item each end point is connected to at least $k$ gateways,
		\item the sum of weights for edges for each gateway $v$ is at most 
		$c$, 
		\begin{equation}
		\sum_{w\in V, \{w,v\}\in E_H} C_{vw} \leq c.
		\end{equation} 
	\end{enumerate}
\end{problem}

The last conditions assures that the redundancy requirement is fulfilled, and 
even if for given station $k-1$ neighbouring gateway would be 
unavailable, it is promised that $k$th gateway will have enough resources for 
serving the station.

\subsection{Greedy algorithm}
\newcommand{\AND}{\textbf{and}\xspace}
\newcommand{\FALSE}{\textbf{false}\xspace}
\newcommand{\TRUE}{\textbf{true}\xspace}

\begin{algorithm}[t!]
	\caption{\label{algorithm:constructive-connection-graph} Greedy algorithm for finding $k$-dominating scheme and its connection scheme. The stop condition is that every vertex is either a gateway or is served by $k$ gateways. If the stop condition is not fulfilled, new gateway according to \textsc{new\_gateway} is chosen. Here
		$G$ is input graph, $C$ is the cost function, $k$ is the domination parameter and $c$ is the common capacity for all gateway nodes}	
	\begin{algorithmic}[1]
		\Function {create\_connection\_graph}{$G=(V,E), C, k, c$}
		\State $E_H\gets\emptyset$
		\State $D \gets \emptyset$
		\State $H \coloneqq (V, E_H, f)$ 
		\While {$\{v\in V: \deg_{H}(v)<k\} \setminus D \neq \emptyset$  }
		\State $w, S \gets \textsc{new\_gateway}(G, H, C, D, k, c)$
		\State $D \gets D \cup \{w\} $.
		\ForAll {$v \in S$}
		\State $E_H \gets E_H\cup \{\{v,w\}\}$
		\EndFor
		\EndWhile
		\State \Return $D,H$
		\EndFunction
	\end{algorithmic}
\end{algorithm}

\begin{algorithm}[t!]
	\caption{\label{algorithm:base-station-choice} The algorithm which chooses best end point to be new gateway. It is chosen to be the end point, which can serve maximal possible stations. Here
		$G$ is input graph, $H$ is the subgraph of $G$, $C$ is the cost function, $D$ is the collection of gateways, $k$ is the domination parameter and $c$ is the common capacity for all gateway nodes. A degree of vertex $v$ in graph $H$ is denoted as $\deg_H(v)$.}	
	\begin{algorithmic}[1]
		\Function {new\_gateway}{$G= (V,E), H=(V,E_H), C, D, k, c$}
		\For{$w \in \rm V$}
		\If{$w\in D$}
		\State ${\rm value}[w]\gets 0$		
		\Else
		\State $T\gets N_G(w)$ \Comment{$N_G(w)$ returns neighbouring 
			nodes}
		\State $T \gets T\setminus D$ \Comment{remove gateways}
		\State $T \gets \{v \in T: \deg_H(v) < k \}$ \Comment{remove served nodes}
		\State $S\gets \emptyset$
		\While{$\sum_{v\in S} C_{vw} \leq c$} \Comment{check if the 
			station can 
			still serve more stations}
		\State $v\gets \arg\min_{v\in T} C_{vw}$ \Comment{choose 
			neighbour with smallest connection cost}
		\State $S \gets S \cup \{v\}$
		\State $T \gets T \setminus\{v\}$
		\EndWhile
		\State ${\rm value}[w] \gets 1 +|S|$ \Comment{because marking $w$ makes it served}
		\EndIf
		\EndFor	
		\State $w \gets \arg\max_{v\in V} {\rm value}[v]$
		\State \Return $w, S$
		\EndFunction
	\end{algorithmic}
\end{algorithm}

Let $G=(V,E)$ be an undirected graph with $n=|V|$ vertices. The algorithm is an extension of the greedy algorithm for $k$-dominating set creation \cite{foerster2013approximating}. It is a greedy algorithm that starts with $D=\emptyset$, and adds a new vertex $w$ minimizing current value of $a(G,D\cup \{w\})$ of the form
\begin{equation}
a(G,D) = nk - \sum_{v\in V}d_k(v,D), \label{eq:preference-foerster}
\end{equation}
where
\begin{equation}
d_k(v,D) = \begin{cases}
\min(k, N(v)\cap D), & v\not \in D,\\
k, &v\in D,
\end{cases}
\end{equation}
where $N(v)$ is the neighbourhood of $v$ in $G$.
The algorithm stops when $a(G,D)=0$, which implies that $D$ is a $k$-dominating set.
Note that function $a(G,D)$ describes how many nodes are already in $D$, or in the case of non-gateways how many gateways are within their radius. 

In our scenario the gateways, which are elements of a $k$-dominating set, 
have 
limited capacity. Thus, we need to store which end points connect to 
newly added gateway. Hence, we start with empty gateway collection $D$ 
and empty graph 
$H=(V,E_H=\emptyset)$.

We choose gateways until $(D,H)$ is a proper solution of our problem. The 
procedure to provide the solution is provided in 
Algorithm~\ref{algorithm:constructive-connection-graph}. At each step, new gateway is chosen to be the one maximizing the number of end point stations that 
it can serve and that requires service~$S$. The details are provided in 
Algorithm~\ref{algorithm:base-station-choice}. When the gateway $w$ is 
chosen, 
it is added to current gateway collection $D=D\cup \{w\}$, and the edges 
set $E_H$ is updated with edges connecting $w$ and nodes it promised to serve 
by $E_H=E_H\cup\{\{w,v\}: v\in S\}$.

\section{Performance evaluation}

The proposed algorithm has been evaluated using the randomly generated topologies and the real topologies of smart metering networks in two Polish cities which datasets we have access to, but are not in public domain. 
The algorithms are compared primarily in terms of the number of base stations selected for a given network topology, method operation time and spreading factors in the communication between selected base stations and end devices assigned to them by each of two algorithms.
Experiments have been carried out for algorithm parameter sample values such as $k \in \{1, 2, 3\}$ and fixed maximal capacity of the gateway. 
The value of the maximal capacity parameter has been determined on the basis of our calculations taking into account transmission parameters such as the probability of packet delivery, the transmission time of a single packet and the transmission window. 

As a reference method, we've used an implementation of the algorithm from~\cite{zhou2019gateway}, as the most similar to our approach. To be more precise, we've applied the heuristic method named accelerated dynamic weighted greedy algorithm (ADWGA) which is algorithm edition based on the recommendations of the authors of \cite{zhou2019gateway} this version of the algorithm for large--scale cases. 
In this approach, the base stations are not limited to the maximum number of end devices that can be assigned to a single base station.
The authors of~\cite{zhou2019gateway} do not provide exacts values of input algorithm parameters, marked in  \cite{zhou2019gateway} as $\alpha$ and $\beta$, but only appoint the ranges within which the parameter values should be. 
Searching for the reference method parameter values has a significant impact on the algorithm's operation time.
Due to the long--running calculations of the reference method, 
we conducted the experiments using the reference method \cite{zhou2019gateway} for two real--life network topologies and for some small random data sets. The results of both methods are compared with each other in table \ref{tab:real-city-table} (for real network topologies) and table \ref{tab:dwga-greedy-results} (for random network topologies).

\begin{figure}[htb]
\centering
	\begin{subfigure}[b]{0.475\textwidth}
		\fbox{\begin{minipage}{0.8\textwidth}
				\includegraphics[clip,width=\textwidth]{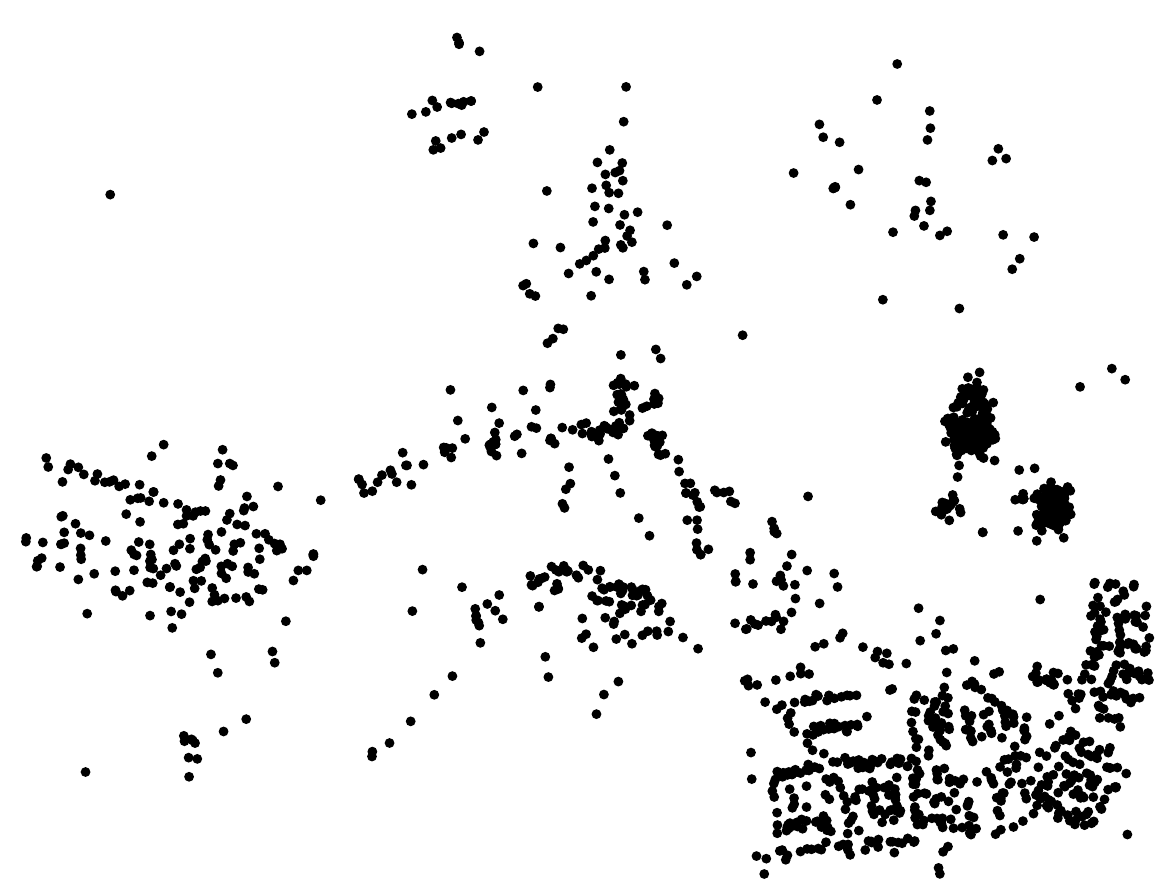}	
		\end{minipage}}
		\caption{City A}
		\label{fig:cityA}
	\end{subfigure}
	\begin{subfigure}[b]{0.475\textwidth}
		\fbox{\begin{minipage}{0.8\textwidth}
				\includegraphics[clip,width=\textwidth]{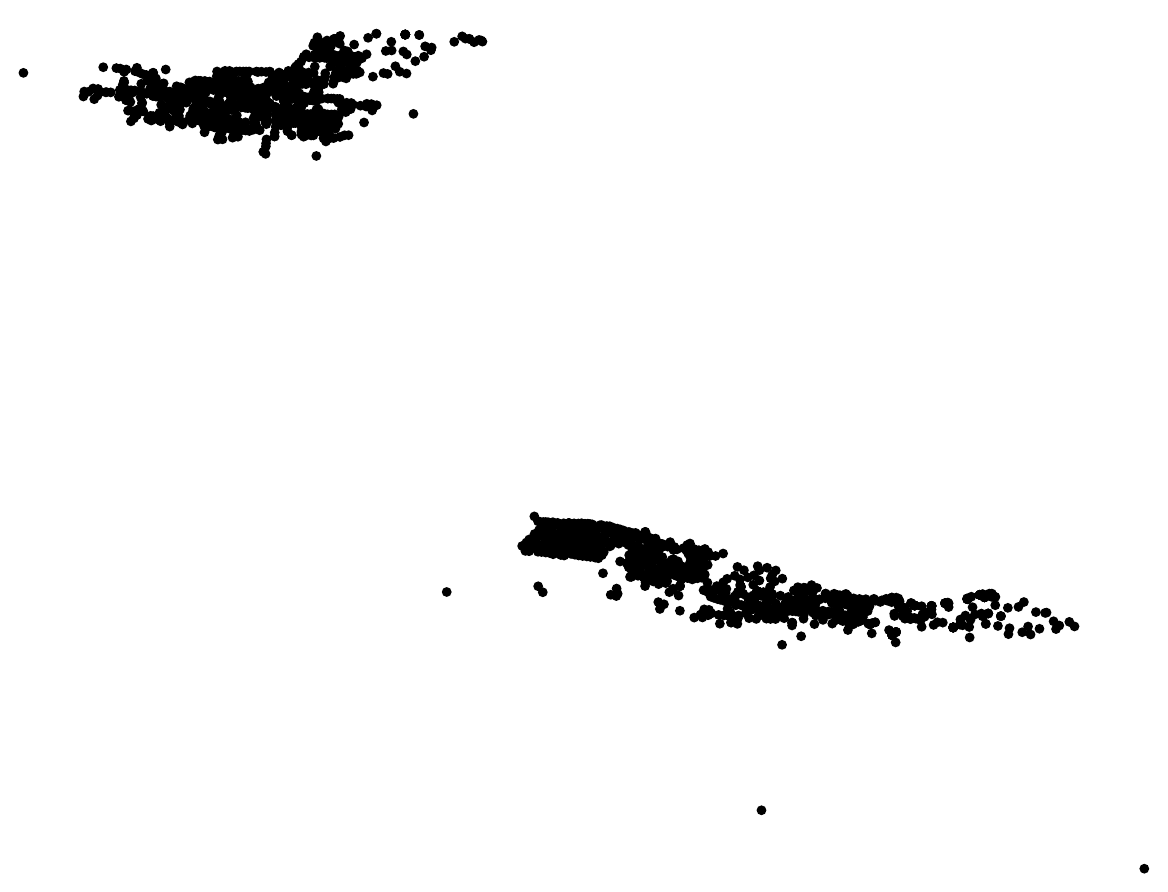}
		\end{minipage}}
		\caption{City B}
		\label{fig:cityB}
	\end{subfigure}
	\caption{Real--life IoT deployment network topologies of two sample cities A and B.}
	\label{fig:real-city-figure}
\end{figure}

Figure \ref{fig:real-city-figure} depicts the distribution of nodes for two real--life IoT deployment topologies of two sample cities A and B, being the subject of experiments. City A is a typical small city topology and city B consists of nodes divided in two separate regions -- a double city estate topology.
 
In table \ref{tab:real-city-table} the results of both greedy and reference method~\cite{zhou2019gateway} for real network topologies are summarized.
The results in the table \ref{tab:real-city-table} include an amount of selected gateways (`number of gateways'), 
algorithm operation time in which the result was obtained (`time [s]') and the average spreading factor to the gateway (`avg. SF'). 
Even though in most cases the reference algorithm found a smaller number of base stations than our method, the difference is very small. 
However, the difference in the time of operation of both methods is huge.
For these two real network topologies our algorithm finds comparable number of base stations
in much less time than reference method.
\begin{table}[htb]

	\resizebox{\textwidth}{!}{
		\begin{tabular}[width=\textwidth]{|l|l|l|l|l|l|l|l|l|l|}
			\hline
			\multicolumn{1}{|c|}{\multirow{2}{*}{city}} & \multicolumn{1}{c|}{\multirow{2}{*}{nodes}} & \multicolumn{1}{c|}{\multirow{2}{*}{\begin{tabular}[c]{@{}c@{}}area\\ {[}m{]}\end{tabular}}} & \multicolumn{1}{c|}{\multirow{2}{*}{k}} & \multicolumn{2}{c|}{number of gateways} & \multicolumn{2}{c|}{\begin{tabular}[c]{@{}c@{}}time \\ {[}s{]}\end{tabular}} & \multicolumn{2}{c|}{avg. SF} \\ \cline{5-10} 
			\multicolumn{1}{|c|}{} & \multicolumn{1}{c|}{} & \multicolumn{1}{c|}{} & \multicolumn{1}{c|}{} & \multicolumn{1}{c|}{greedy} & \multicolumn{1}{c|}{reference \cite{zhou2019gateway}} & \multicolumn{1}{c|}{greedy} & \multicolumn{1}{c|}{reference \cite{zhou2019gateway}} & \multicolumn{1}{c|}{greedy} & \multicolumn{1}{c|}{reference \cite{zhou2019gateway}} \\ \hline
			\multirow{3}{*}{A} & \multirow{3}{*}{1719} & \multirow{3}{*}{2490 $\times$ 4090} & 1 & 6 & 6 & 4.95 & 253.08 & 9.19 & 9.18 \\ 
			&  &  & 2 & 12 & 11 & 8.86 & 3118.27 & 9.2 & 8.92 \\ 
			&  &  & 3 & 17 & 15 & 12.58 & 7603.15 & 9.17 & 8.88 \\ \hline
			\multirow{3}{*}{B} & \multirow{3}{*}{1760} & \multirow{3}{*}{1260 $\times$ 3060} & 1 & 8 & 7 & 2.47 & 169.45 & 7.8 & 8.06 \\ 
			&  &  & 2 & 15 & 14 & 4.57 & 363.16 & 7.84 & 7.86 \\ 
			&  &  & 3 & 20 & 19 & 6.72 & 2249.87 & 7.84 & 7.89\\ \hline
	\end{tabular}}
	\caption{\label{tab:real-city-table}Results of the proposed greedy algorithm and reference method \cite{zhou2019gateway} (ADWGA) for two real--life network topologies~A~and~B.}
\end{table}

Experimental results presented below include proposed algorithm performance statistics for various scenarios assuming the different number of network nodes randomly positioned in accordance with the uniform distribution in changing areas. The algorithm has been tested on network topologies with different density of nodes distribution in space.
The statistics presented in tables below were obtained based on 30 runs of the algorithm for each number of nodes equals 1000, 2500, 5000, 10000 and 20000 distributed in a rectangle with dimensions 5000m~$\times$~7500m, 10000m~$\times$~15000m, 15000m~$\times$~22500m and 20000m~$\times$~30000m. Before each running of the algorithm positions of nodes were selected randomly. Tables ~\ref{tab:stat-k1}, \ref{tab:stat-k2} and \ref{tab:stat-k3} contain statistics such as the average number of selected gateways, the average algorithm running time and the average spreading factor occurring between gateways and their clients (end point nodes) for redundancy factor $k \in \{1, 2, 3\}$ respectively and with fixed maximal capacity.

\begin{table}[htb]

	\resizebox{\textwidth}{!}{
		\begin{tabular}[width=\textwidth]{|l|l|l|l|l|l|l|l|l|}
			\hline
			\multirow{2}{*}{nodes} & \multicolumn{1}{c|}{\multirow{2}{*}{\begin{tabular}[c]{@{}c@{}}area\\ {[}m{]}\end{tabular}}} & \multirow{2}{*}{k} & \multicolumn{2}{c|}{\begin{tabular}[c]{@{}c@{}}avg. number of\\  gateways\end{tabular}} & \multicolumn{2}{c|}{\begin{tabular}[c]{@{}c@{}}avg. time\\ {[}s{]}\end{tabular}} & \multicolumn{2}{c|}{avg. SF} \\ \cline{4-9} 
			& \multicolumn{1}{c|}{} &  & greedy & reference \cite{zhou2019gateway} & greedy & reference \cite{zhou2019gateway}& greedy & reference \cite{zhou2019gateway} \\ \hline
			\multirow{3}{*}{1000} & \multirow{3}{*}{5000 $\times$ 7500} & 1 & 16.31 & 15.37 & 0.75 & 71.74 & 9.52 & 9.44 \\ 
			&  & 2 & 29.9 & 27.5 & 1.41 & 640.51 & 9.54 & 9.58 \\ 
			&  & 3 & 42.38 & 38.31 & 2.03 & 1139.58 & 9.57 & 9.56 \\ \hline
			\multirow{3}{*}{1000} & \multirow{3}{*}{10000 $\times$ 15000} & 1 & 47.66 & 47.94 & 0.23 & 64.58 & 9.62 & 9.61 \\ 
			&  & 2 & 86.03 & 83.2 & 0.43 & 733.19 & 9.67 & 9.66 \\ 
			&  & 3 & 123.34 & 117.69 & 0.61 & 933.72 & 9.7 & 9.68 \\ \hline
			\multirow{3}{*}{1000} & \multirow{3}{*}{15000 $\times$ 22500} & 1 & 91.34 & 92.59 & 0.13 & 80.6 & 9.69 & 9.69 \\ 
			&  & 2 & 166.45 & 166.41 & 0.25 & 939.71 & 9.74 & 9.71 \\ 
			&  & 3 & 236.34 & 249.41 & 0.34 & 1074.58 & 9.77 & 9.7 \\ \hline
			\multirow{3}{*}{1000} & \multirow{3}{*}{20000 $\times$ 30000} & 1 & 145.38 & 153.93 & 0.1 & 348.31 & 9.74 & 9.74 \\ 
			&  & 2 & 264.31 & 296.03 & 0.19 & 739.93 & 9.78 & 9.71 \\ 
			&  & 3 & 373.38 & 566.28 & 0.27 & 896.88 & 9.78 & 10.02 \\ \hline
			\multirow{3}{*}{2500} & \multirow{3}{*}{5000 $\times$ 7500} & 1 & 18.14 & 16.59 & 5.4 & 624.57 & 9.46 & 9.38 \\ 
			&  & 2 & 32.52 & 30.24 & 9.94 & 8496.08 & 9.48 & 9.52 \\ 
			&  & 2 & 46.72 & 41.14 & 14.39 & 18170.28 & 9.5 & 9.5 \\ \hline
	\end{tabular}}
	\caption{\label{tab:dwga-greedy-results}Results of the greedy and reference \cite{zhou2019gateway} method for sample topology scenarios.}
\end{table}

In table \ref{tab:dwga-greedy-results} greedy and reference method statistics, described above, for small data sets are contained. 
It can be seen that in each case the running time of the greedy algorithm is much shorter than the running time of the reference method, 
especially for network topologies characterized by a dense distribution of nodes in space.
In cases where the reference algorithm returns a smaller number of base stations than our method, which is on the plus side for the ADWGA, because it generates lower costs, then the greedy algorithm returns a little more base stations in a much shorter time. Moreover, for less dense network topologies our algorithm finds significantly fewer base stations much faster than reference method. 
The speed of the greedy algorithm allows us to obtain results for large--scale network topologies in a relatively short time.

\begin{figure}[htb]
	\centering
	\begin{subfigure}{0.5\textwidth}
		\includegraphics[width=0.98\textwidth]{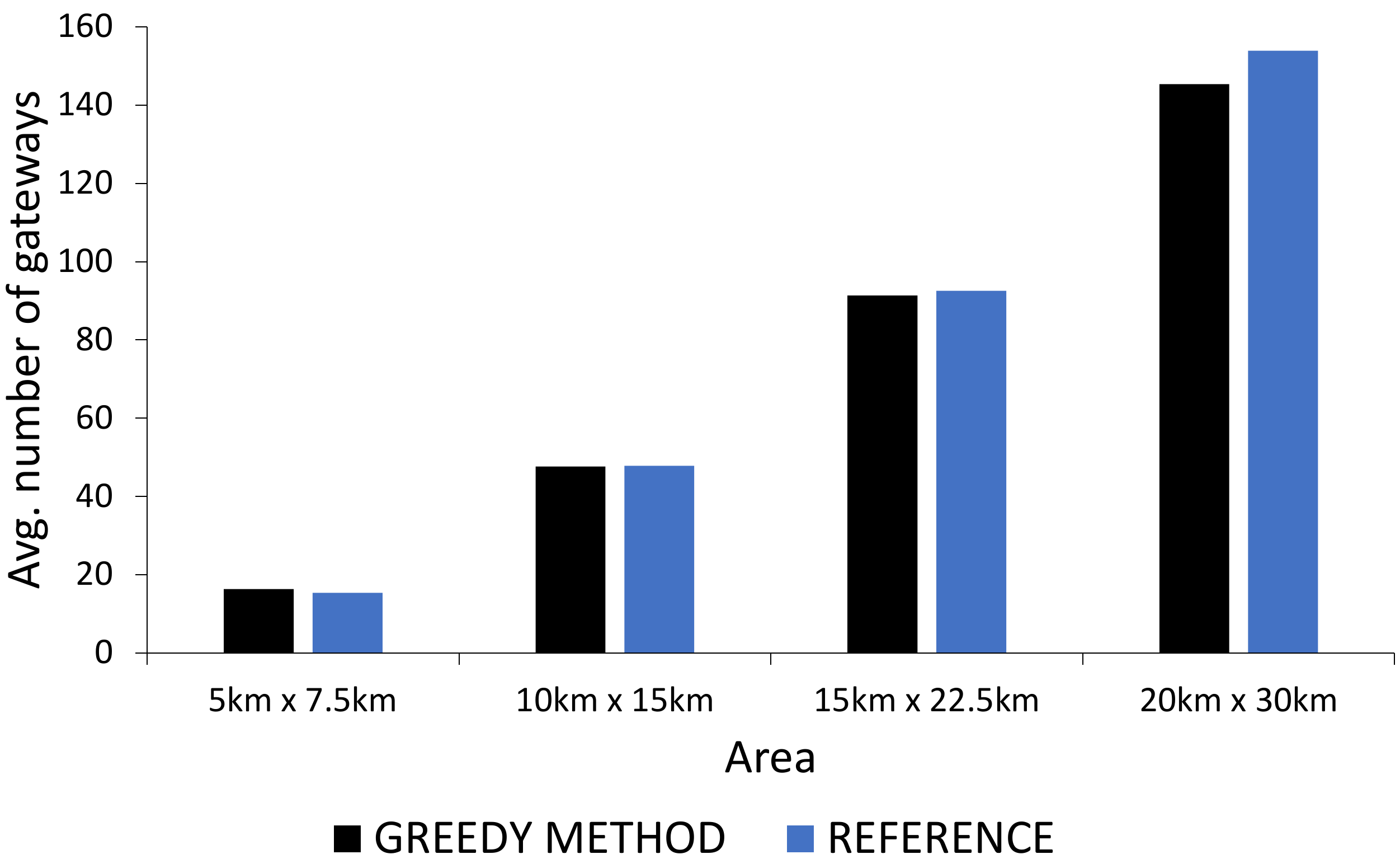}
		\caption{1000 nodes, k=1}
		\includegraphics[width=\textwidth]{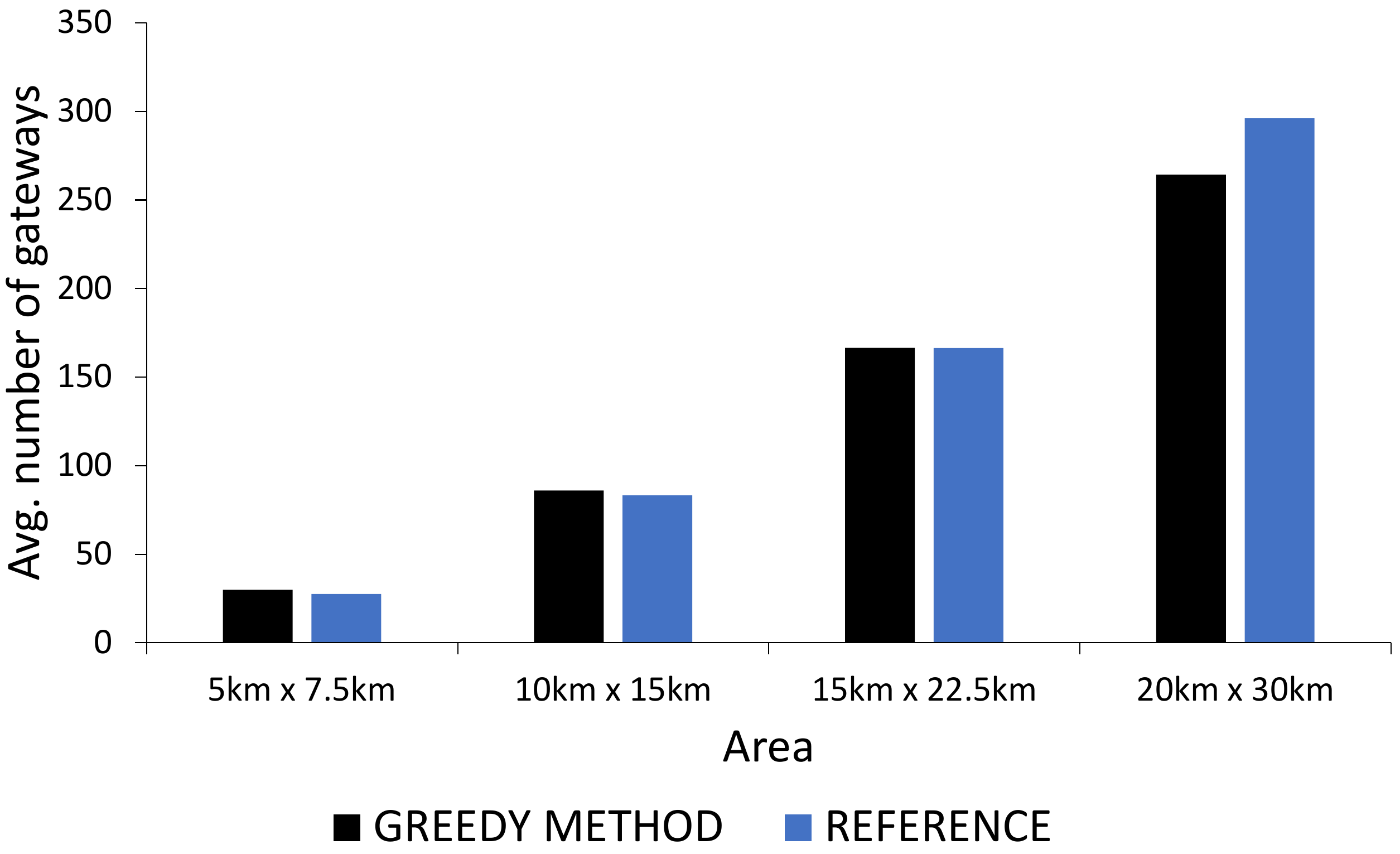}
		\caption{1000 nodes, k=2}
	\end{subfigure}%
	\begin{subfigure}{0.48\textwidth}
		\includegraphics[width=\textwidth]{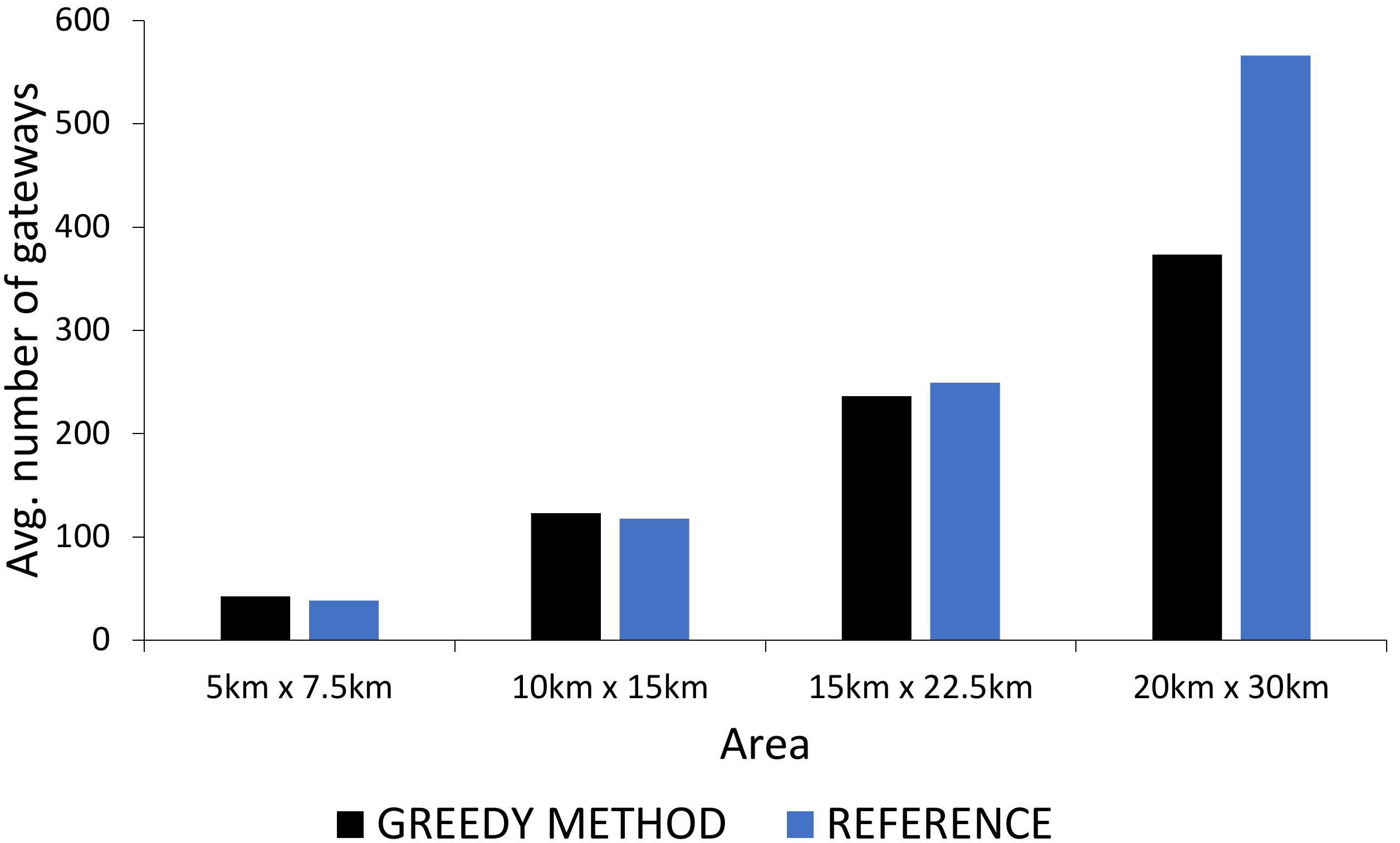}
		\caption{1000 nodes, k=3}
		\includegraphics[width=\textwidth]{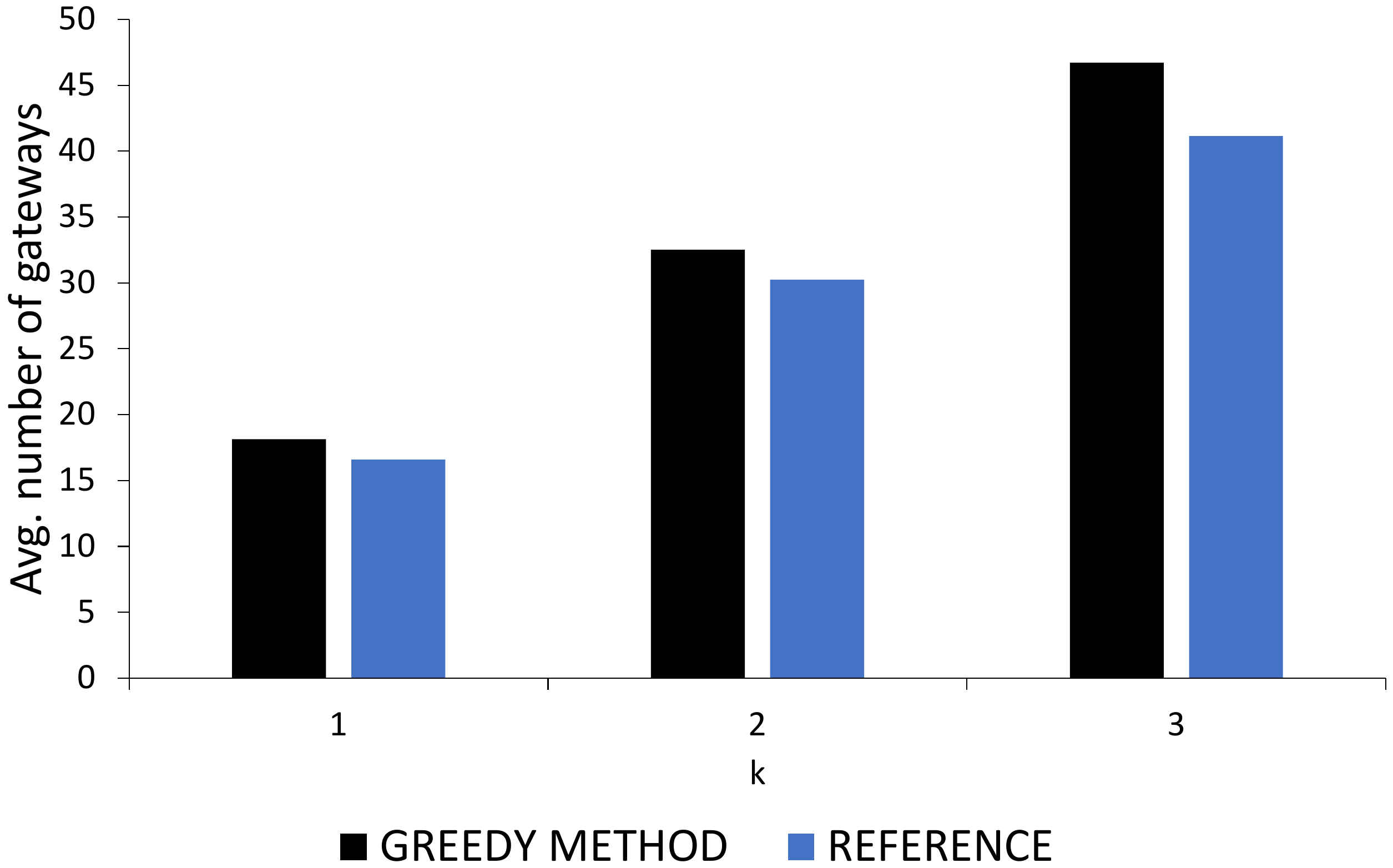}
		\caption{2500 nodes, $k \in \{1, 2, 3\}$}
	\end{subfigure}
	\caption{Comparison of the average number of selected gateways for $k \in \{1, 2, 3\}$ for both methods; the proposed greedy algorithm and reference method \cite{zhou2019gateway} for 1000 number of nodes deployed in all areas (a), (b), (c) and for 2500 number of nodes deployed in the  smallest of considered areas (d).}
	\label{fig:adwga-greedy-gateways}
\end{figure}

Figure \ref{fig:adwga-greedy-gateways} is the graphic interpretation of results for the average number of selected gateways presented in table \ref{tab:dwga-greedy-results}.
To make the charts more transparent, the dimensions of the network topology area are presented in units of kilometers instead of meters, as in the tables.

\begin{table}[htb]
	\centering

	\resizebox{0.7\textwidth}{!}{
		\noindent\begin{tabular}{|l|l|l|l|l|l|}\hline
			\multicolumn{1}{|c|}{nodes} & \multicolumn{1}{c|}{area}  & \multicolumn{1}{c|}{avg. number} &	\multicolumn{1}{c|}{avg. time} & \multicolumn{1}{c|}{avg. SF}\\
			& \multicolumn{1}{c|}{[m]} & \multicolumn{1}{c|}{of gateways} &	\multicolumn{1}{c|}{[s]} & \\\hline
			2500 &	10000 $\times$ 15000	&	53.9  &	1.97 & 9.54\\
			&	15000 $\times$ 22500	&	105.9 &	0.97 & 9.64\\
			&	20000 $\times$ 30000	&	170.6 &	0.61 & 9.7\\\hline
			5000 &	5000 $\times$ 7500	    &	19.53 & 34.79 & 9.43\\
			&	10000 $\times$ 15000	&	58.73 &	8.74 & 9.49\\
			&	15000 $\times$ 22500	&	115.53 & 4.06 & 9.57\\
			&  20000 $\times$ 30000	&	188.03 & 2.51 & 9.64\\\hline
			10000 &	5000 $\times$ 7500		&   20.97 & 155.8 &9.41\\
			&	10000 $\times$ 15000	&	62.53 &	36.39 & 9.45\\
			&	15000 $\times$ 22500	&	124.4 &	15.07 & 9.52\\
			&	20000 $\times$ 30000	&	203.8 &	8.78 & 9.58\\\hline
			20000 &  5000 $\times$ 7500		&   22.73  & 1015.3 & 9.21\\
			&	10000 $\times$ 15000	&	67.1  & 177.7 & 9.43\\
			&	15000 $\times$ 22500	&	132.7 & 72.3 & 9.47\\
			&	20000 $\times$ 30000	&	218.9 & 36.3 & 9.52\\\hline		 
	\end{tabular}}
	\caption{\label{tab:stat-k1}Statistics of the proposed greedy algorithm of finding $k$ -- dominating set for $k=1$.}
\end{table}

\begin{table}[htb]

	\centering
	\resizebox{0.7\textwidth}{!}{%
		\begin{tabular}[H]{|l|l|l|l|l|l|}\hline
			\multicolumn{1}{|c|}{nodes} & \multicolumn{1}{c|}{area}  & \multicolumn{1}{c|}{avg. number} &	\multicolumn{1}{c|}{avg. time} & \multicolumn{1}{c|}{avg. SF}\\
			& \multicolumn{1}{c|}{[m]} & \multicolumn{1}{c|}{of gateways} &	\multicolumn{1}{c|}{[s]} & \\\hline
			2500 &	10000 $\times$ 15000	&	97.1	& 3.41 & 9.61\\
			&	15000 $\times$ 22500	&	188.43	& 1.60 & 9.7\\
			&	20000 $\times$ 30000	&	305.07	& 1.08 & 9.75\\\hline
			5000  &	5000 $\times$ 7500	    &	35.43	& 59.42 & 9.48\\
			&	10000 $\times$ 15000	&	104.80	& 14.84 & 9.55\\
			&	15000 $\times$ 22500	&	206.03	& 7.07 & 9.65\\
			&	20000 $\times$ 30000	&	333.27	& 4.31 & 9.71\\\hline	  
			10000 &	5000 $\times$ 7500m	    &	37.57   & 264.52 & 9.45\\
			&	10000 $\times$ 15000	&	111.63	& 61.24 & 9.52\\
			&	15000 $\times$ 22500	&	220.3	& 25.55 & 9.59\\
			&	20000 $\times$ 30000	&	361.53	& 14.91 & 9.65\\\hline
			20000 &	5000 $\times$ 7500m	    &   40.83	& 1735.58 & 9.25\\
			&	10000 $\times$ 15000	&	117.23	& 301.56 & 9.48\\
			&	15000 $\times$ 22500	&	233.81	& 121.46 & 9.54\\
			&	20000 $\times$ 30000	&	385.35	& 61.6 & 9.59\\\hline	  		 
	\end{tabular}}
	\caption{\label{tab:stat-k2} Statistics of the proposed greedy algorithm of finding $k$ -- dominating set for $k=2$.}
\end{table}

\begin{table}[htb]
	\caption{\label{tab:stat-k3}Statistics of the proposed greedy algorithm of finding $k$ -- dominating set for $k=3$.}
	\centering
	\resizebox{0.7\textwidth}{!}{%
		\begin{tabular}[H]{|l|l|l|l|l|l|}\hline
			\multicolumn{1}{|c|}{nodes} & \multicolumn{1}{c|}{area}  & \multicolumn{1}{c|}{avg. number} &	\multicolumn{1}{c|}{avg. time} & \multicolumn{1}{c|}{avg. SF}\\
			& \multicolumn{1}{c|}{[m]} & \multicolumn{1}{c|}{of gateways} &	\multicolumn{1}{c|}{[s]} & \\\hline
			2500 & 10000  $\times$ 15000	&	137.43	& 4.62 & 9.58\\
			& 15000  $\times$ 22500	&	266.5	& 2.21 & 9.68\\
			& 20000  $\times$ 30000	&	430.87	& 1.54 & 9.74\\\hline
			5000    & 5000 $\times$ 7500	&   50.27	& 81.38 & 9.49\\
			& 10000 $\times$ 15000	&   147.63  & 20.66 & 9.58\\
			& 15000 $\times$ 22500	&	289.57	& 9.82 & 9.68\\
			& 20000 $\times$ 30000	&	470.13	& 6.07 &9.74\\\hline
			10000	& 5000  $\times$ 7500	&	53.33	& 366.6 & 9.47\\
			& 10000  $\times$ 15000	&	157.63	& 84.68 & 9.54\\
			& 15000  $\times$ 22500	&	309.2	& 35.47 & 9.62\\
			& 20000  $\times$ 30000	&	506.96	& 20.9 & 9.68\\\hline
			20000   & 5000 $\times$ 7500m	&   58.3	& 2532.12 & 9.26\\
			& 10000 $\times$ 15000	&	165.62	& 422.78 & 9.5\\
			& 15000 $\times$ 22500	&	327.15	& 171.19 & 9.57\\
			& 20000 $\times$ 30000	&	539.58	& 86.86 &9.62\\\hline			
		\end{tabular}}
\end{table}

\begin{figure}[H]
	\begin{subfigure}{0.5\textwidth}
		\includegraphics[width=\linewidth]{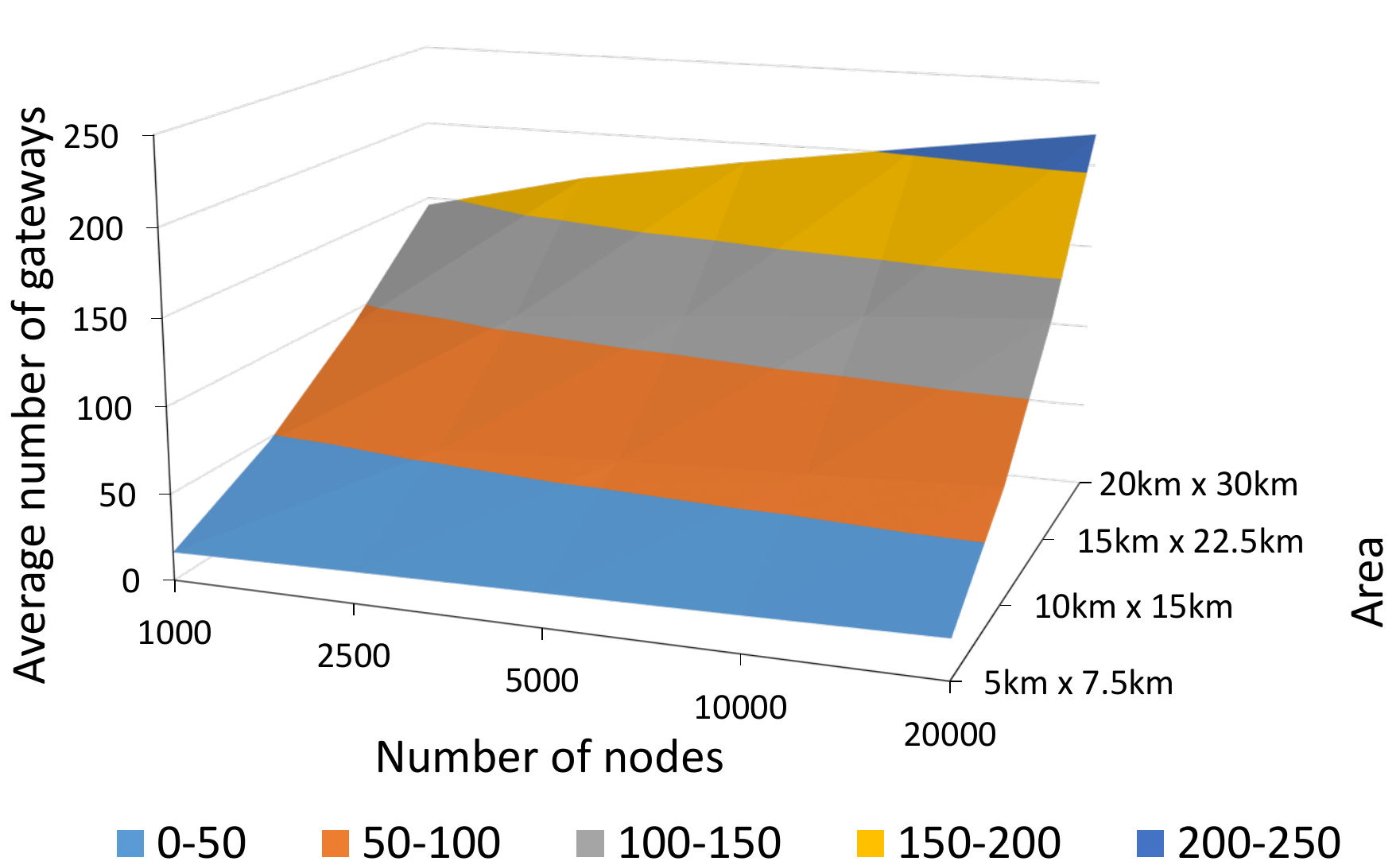}
		\caption{k = 1}
	\end{subfigure}\hspace{0.2cm}
	\begin{subfigure}{0.5\textwidth}
		\includegraphics[width=\linewidth]{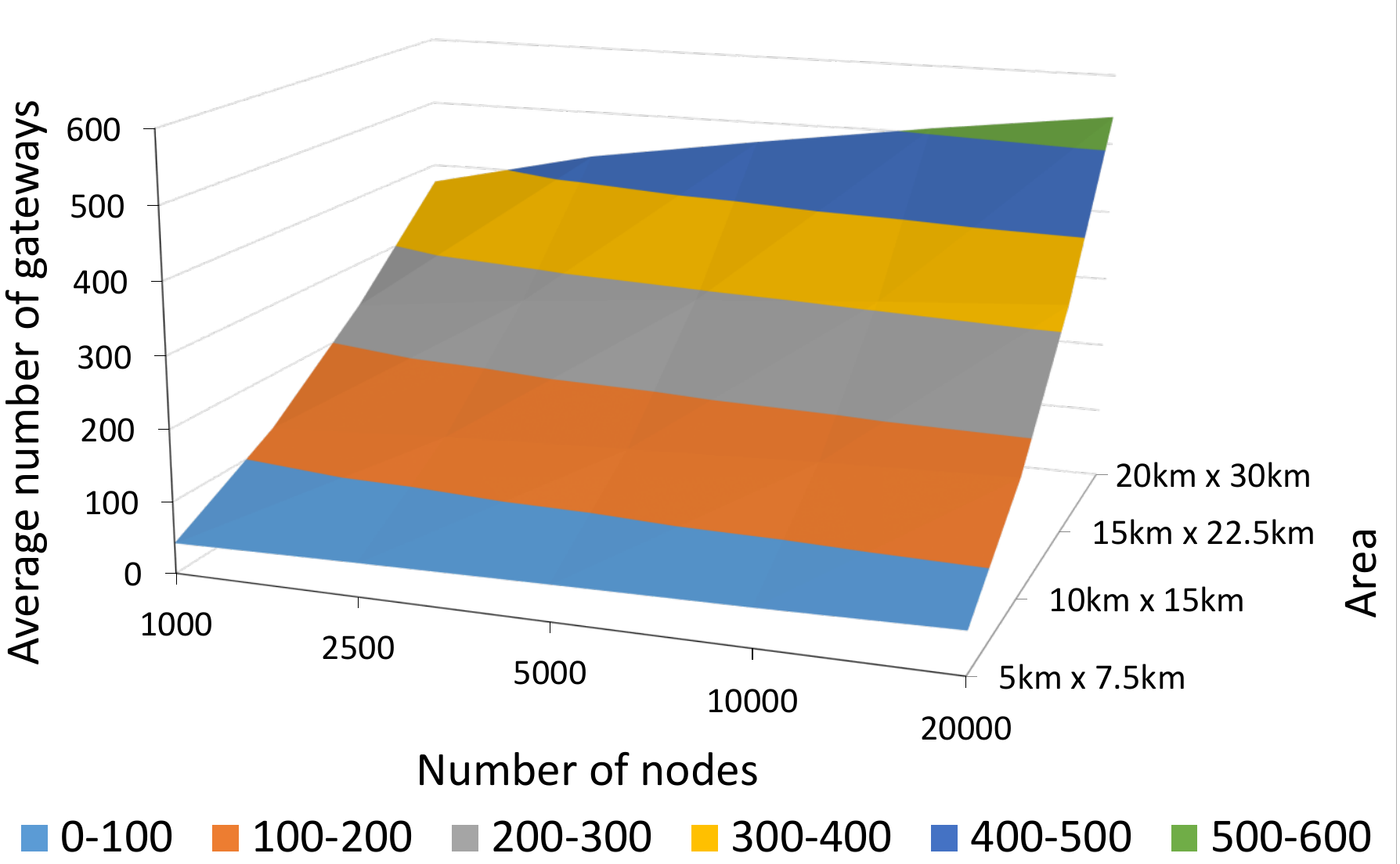}
		\caption{k = 2}
	\end{subfigure}
	\caption{Average number of selected gateways for each scenario for redundancy factor $k \in \{1, 2\}$.}
	\label{fig:avg-nr-gw}
\end{figure}

\begin{figure}[!h]
	\begin{subfigure}{0.5\linewidth}
		\includegraphics[width=\linewidth]{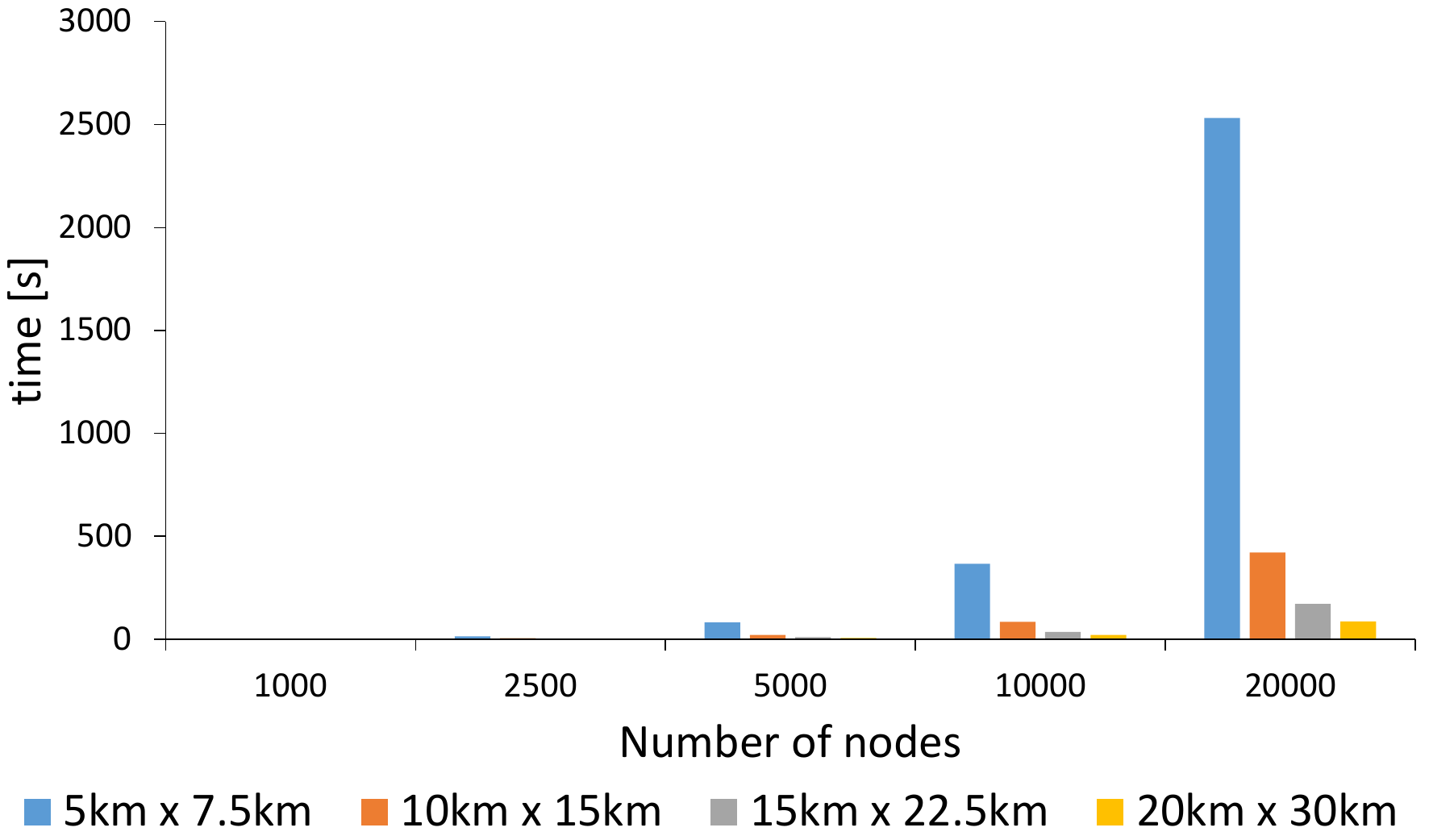}
		\caption{k = 1}
		\label{fig:avg-time-k1}
	\end{subfigure}\hspace{0.2cm}
	\begin{subfigure}{0.5\linewidth}
		\includegraphics[width=\linewidth]{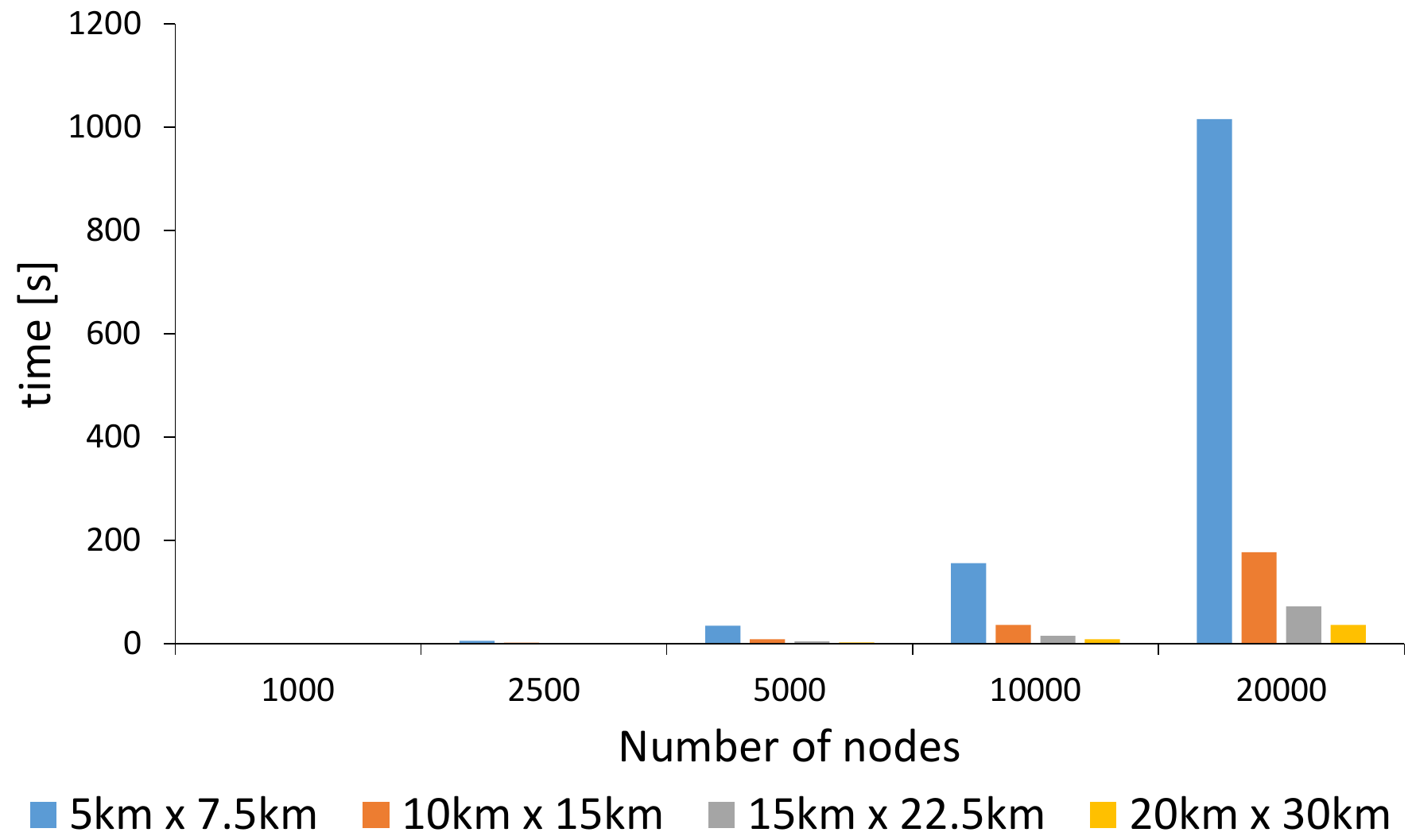}
		\caption{k = 2}
		\label{fig:avg-time-k2}
	\end{subfigure}
	\caption{\label{avg-time} Average greedy algorithm run time for each scenario for redundancy factor $k \in \{1, 2\}$.}
	\label{fig:avg-time}
\end{figure}

\begin{figure}[H]
	\begin{subfigure}{0.5\textwidth}
		\includegraphics[width=\textwidth]{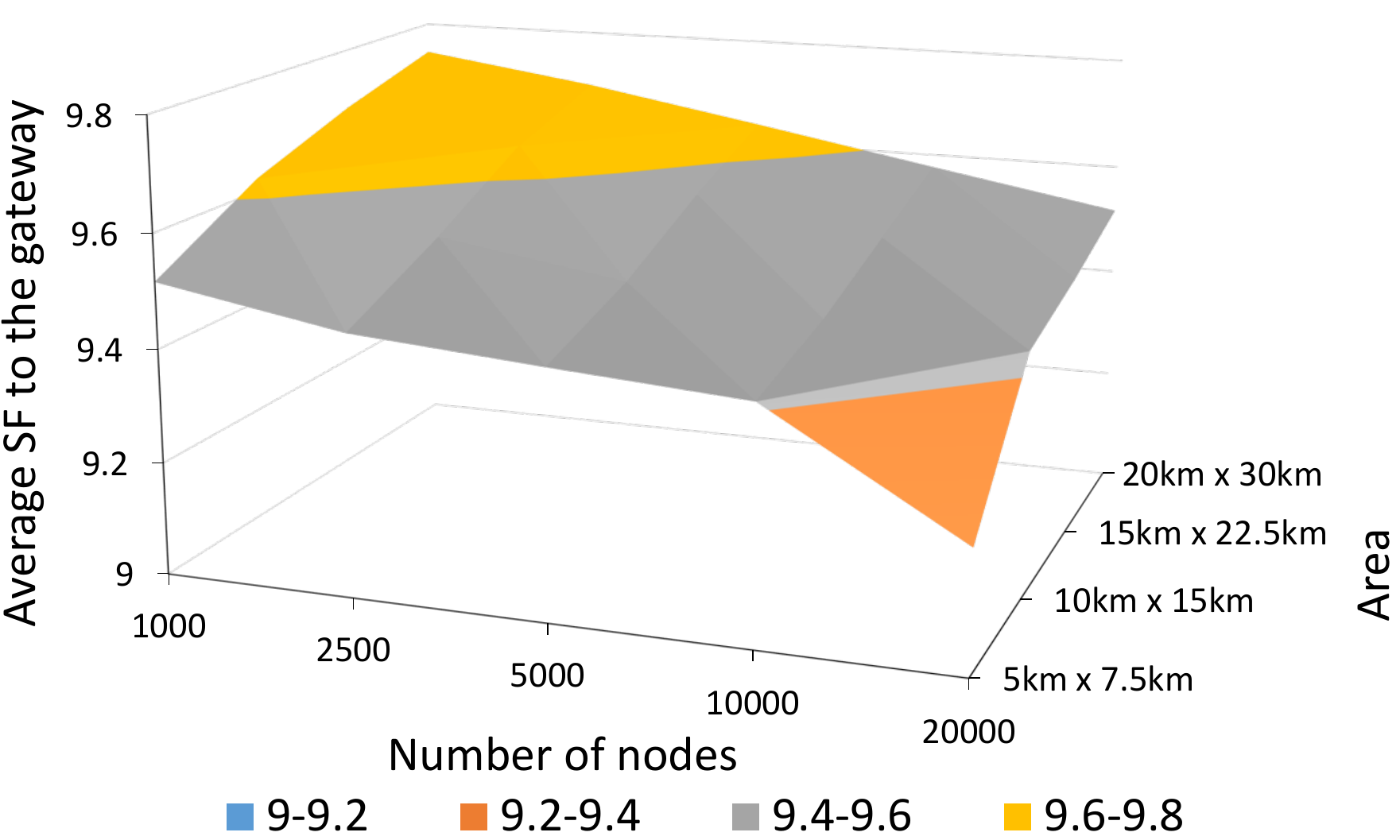}
		\caption{k = 1}
		\label{fig:avg-sf-k1}
	\end{subfigure}\hspace{0.2cm}
	\begin{subfigure}{0.5\textwidth}
		\includegraphics[width=\textwidth]{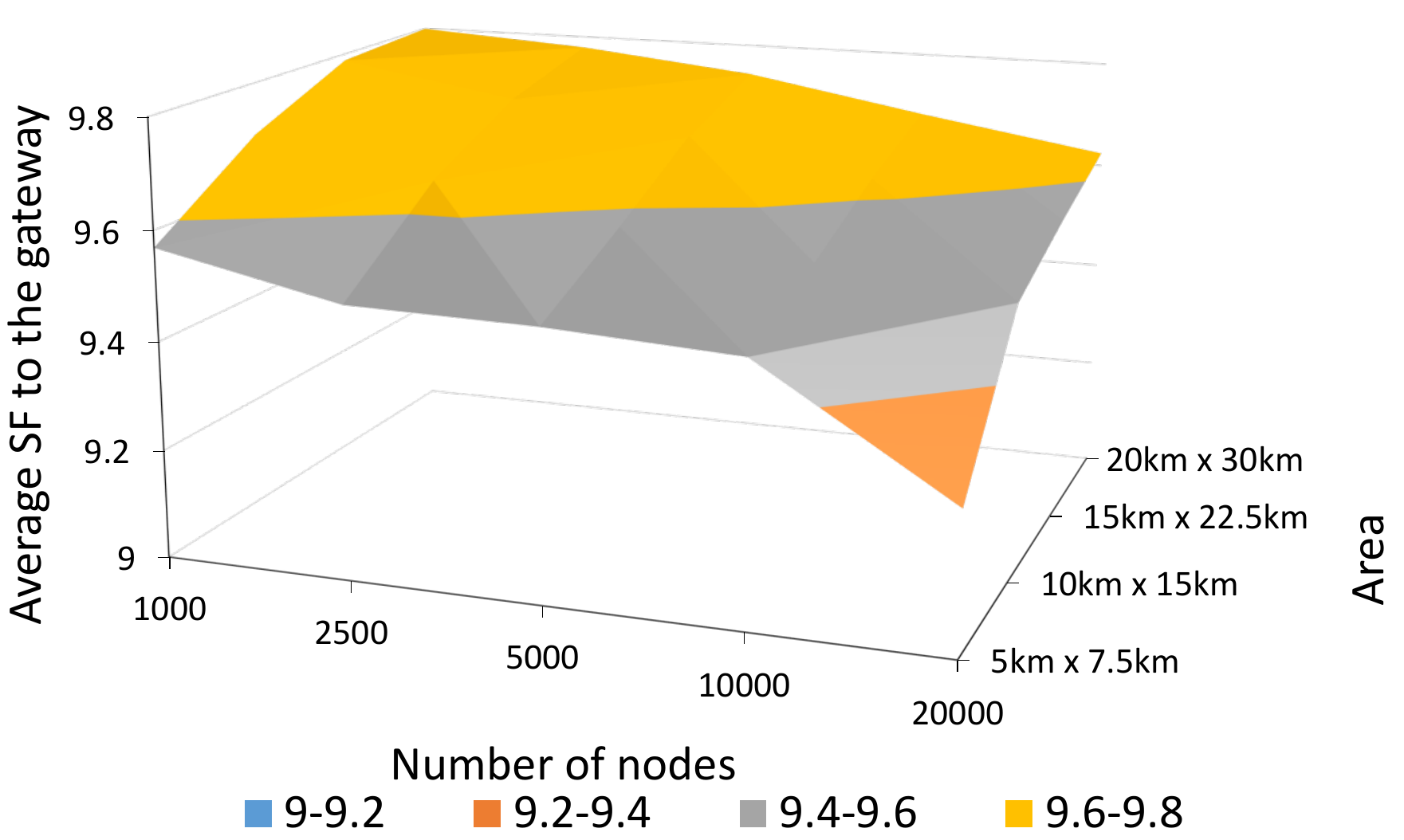}
		\caption{k = 2}
		\label{fig:avg-sf-k2}
	\end{subfigure}
	\caption{\label{avg-sf} Average Spreading Factor to the gateway for each scenario for redundancy factor $k \in \{1, 2\}$.}
	\label{fig:avg-sf}
\end{figure}

On figures \ref{fig:avg-nr-gw}, \ref{fig:avg-time} and \ref{fig:avg-sf} a graphic interpretation of the greedy method has been shown to ilustrate the results presented in the tables \ref{tab:dwga-greedy-results}, \ref{tab:stat-k1}, \ref{tab:stat-k2}, \ref{tab:stat-k3} for $k \in \{1, 2\}$. 

\begin{figure}[tp]
		\centering
	\begin{subfigure}{0.8\textwidth}
		\includegraphics[trim={0.1cm 10.7cm 0.4cm 12cm},clip,width=\linewidth]{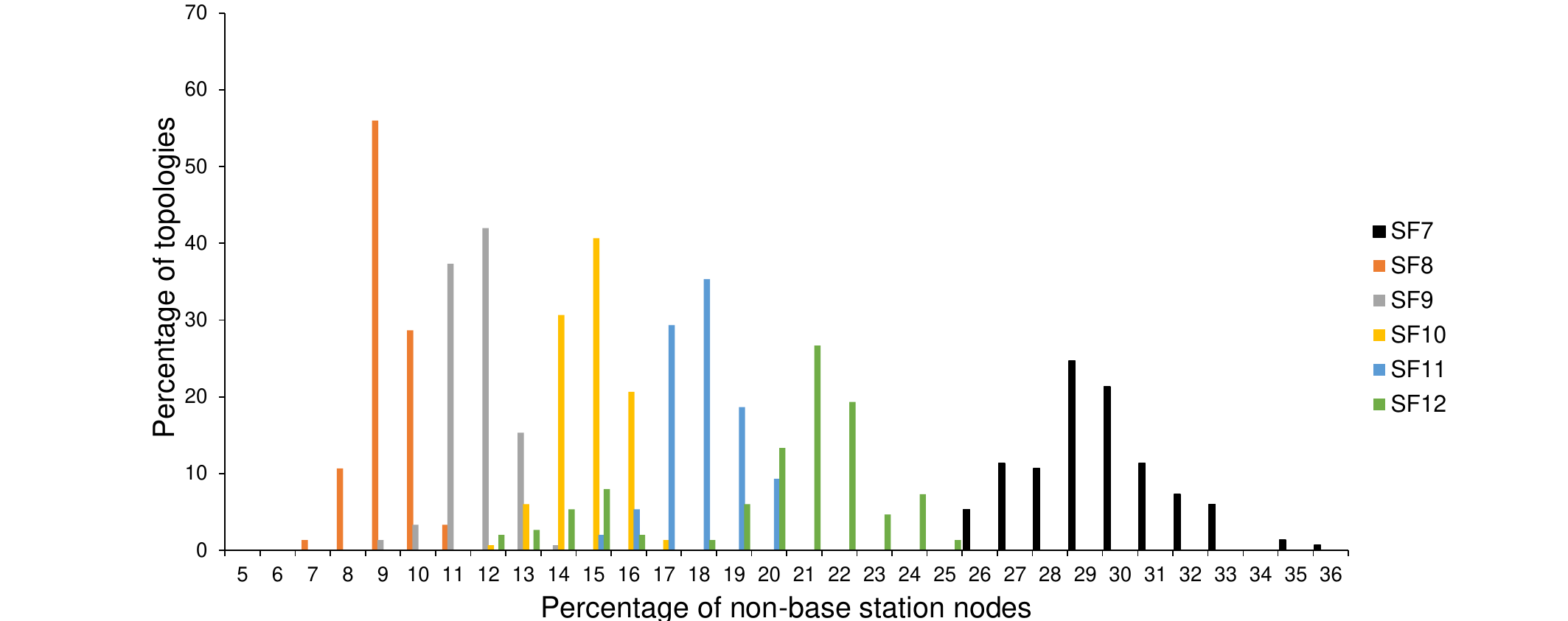}
		\caption{Area: 5000m $\times$ 7500m}
		\label{fig:dist-SF-5000m-k1}
	\end{subfigure}
	\begin{subfigure}{0.8\textwidth}
		\includegraphics[trim={0.1cm 10.7cm 0.4cm 12cm},clip,width=\linewidth]{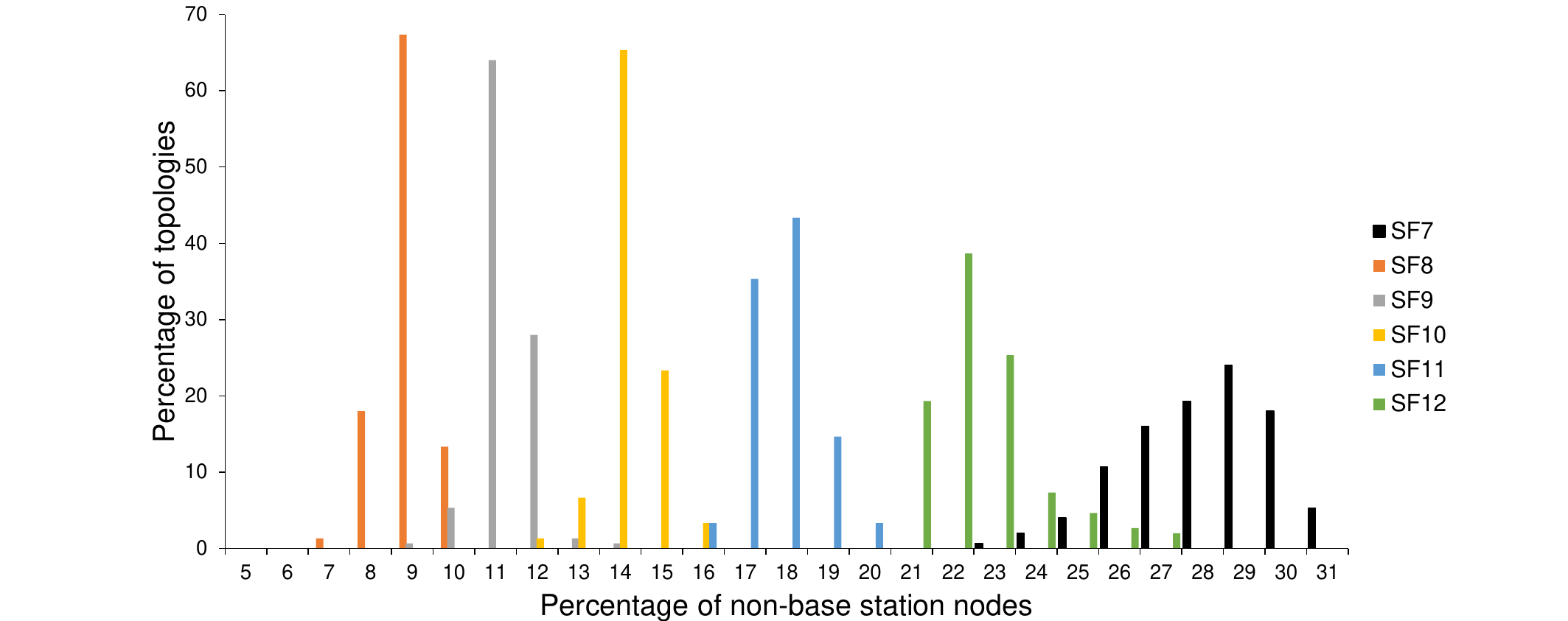}
		\caption{Area: 10000m $\times$ 15000m}
		\label{fig:dist-SF-10000m-k1}
	\end{subfigure}
	\begin{subfigure}{0.8\textwidth}
		\includegraphics[trim={0.1cm 10.7cm 0.4cm 12cm},clip,width=\linewidth]{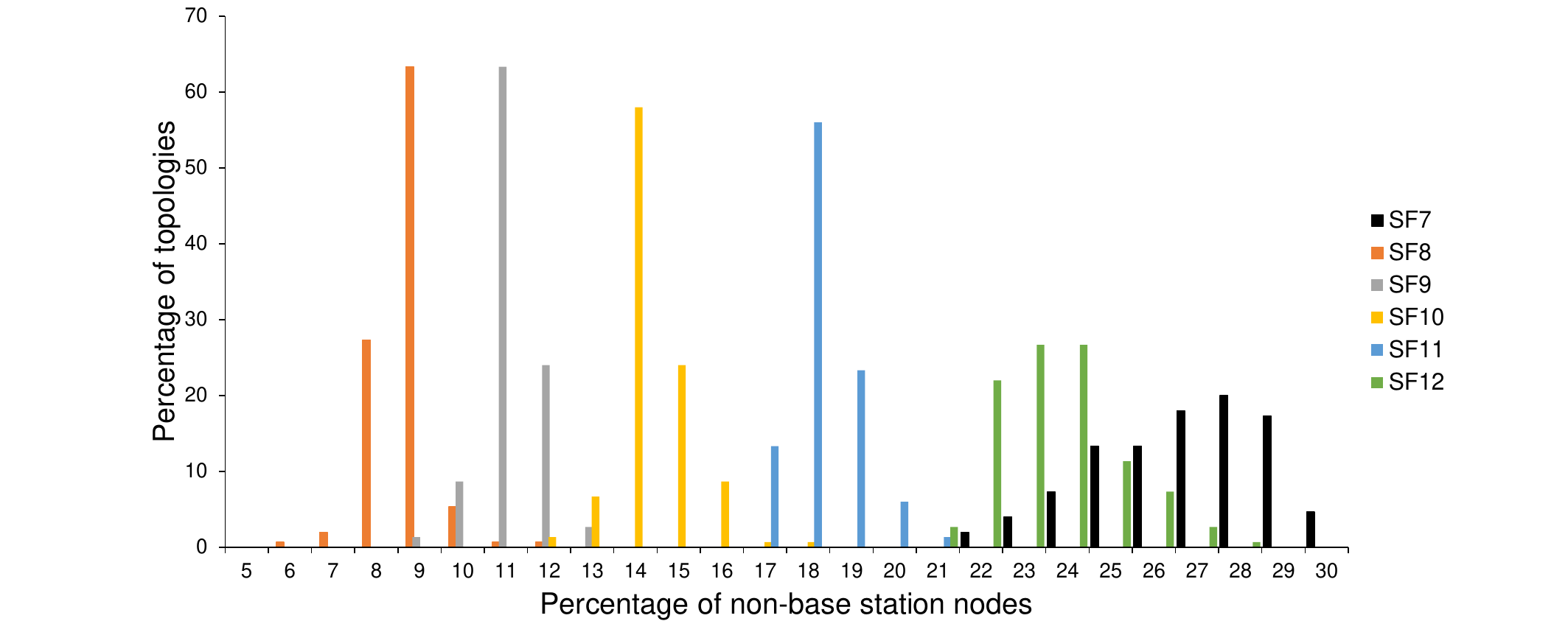}
		\caption{Area: 15000m $\times$ 22500m}
		\label{fig:dist-SF-15000m-k1}
	\end{subfigure}
	\begin{subfigure}{0.8\textwidth}
		\includegraphics[trim={0.1cm 10.7cm 0.4cm 12cm},clip,width=\linewidth]{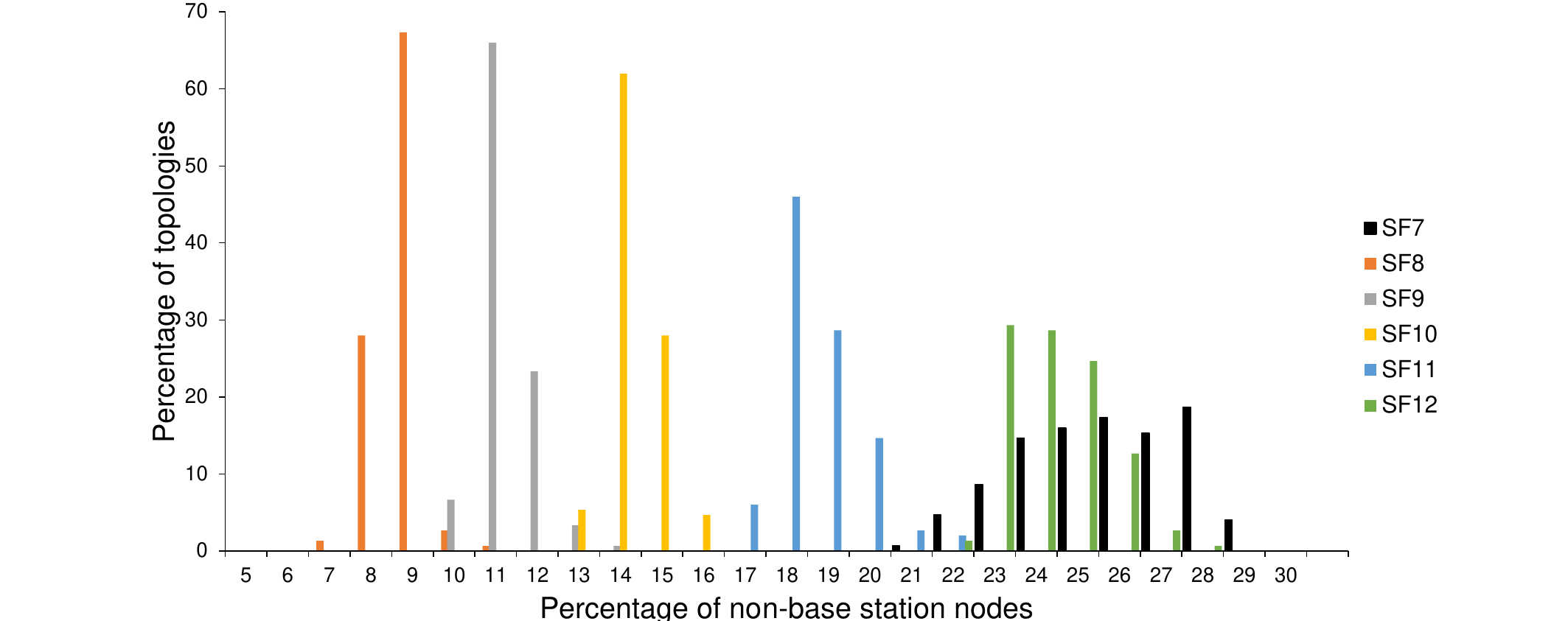}
		\caption{Area: 20000m $\times$ 30000m}
		\label{fig:dist-SF-20000m-k1}
	\end{subfigure}
	\caption{Distributions of the percentage number of nodes having a specific SF to the gateway for different sizes of the node distribution area.}
	\label{SF-distribution}
\end{figure}

The distribution plots shown in the figure \ref{SF-distribution} present 
how the percentage number of non -- gateway having a given spreading factor to the gateway is distributed for a sample consisting of specified network topologies. 
A single sample consists of 150 topologies with an amount of nodes equal to 1000, 2500, 5000, 10000, 20000 distributed in a given area,
where 30 random topologies were generated for each of the five number of nodes. 
Each of the plots \ref{fig:dist-SF-5000m-k1}, \ref{fig:dist-SF-10000m-k1}, \ref{fig:dist-SF-15000m-k1}, \ref{fig:dist-SF-20000m-k1} was created on the basis of described 150 topologies generated for each area separately. 

The distributions of percentage number of non -- gateway nodes having a spreading factor equal to SF8, SF9, SF10 or SF11 to the target gateway are in very similar ranges of values, regardless of the size of the area.
For instance, the percentage of the number of nodes having SF8 to gateway from topologies distributed on area 5000m~$\times$~7500m is in the range from 6\% to 11\% and stays within similar limits despite the increasing area and, consequently, the decreasing network density. The same phenomena can be observed for percentage of number of nodes that have spreading factor to gateway equals to SF9, SF10 and SF11. The algorithm selects the locations of gateway in such a way that the percentage of clients with a given SF to the gateway is very similar, regardless of the density of the network.
Around each gateway, there are areas within there are non -- gateway nodes having a given spreading factor SF7, SF8, etc. The higher the density, the proportionally higher percentage of nodes is in the area with spreading factor SF7 and less in the area with SF12. 
Similarly, the lower the density, the proportionally smaller percentage of nodes falls into the area with spreading factor SF7 to the gateway and more into the area with spreading factor SF12.
In the remaining areas, the number of nodes in each of them is proportionally comparable regardless of the density of the network.
In the figure \ref{SF-distribution} can be observed that as the area increases and, as a consequence, the network density decreases, the values of the bins in distributions of nodes with SF7 to the gateway decrease. In the subsequent charts, the percentage range of nodes having SF7 to the gateway moves left along the horizontal axis, simultaneously the percentage range of nodes with SF12 to the gateway moves right along the same axis.
In other words, the lower the network density, the lower the percentage of nodes having SF7 and higher percentage of nodes with SF12 to the gateway.

\begin{figure}[ht]
\centering
	\begin{subfigure}{.45\textwidth}
		\includegraphics[width=\linewidth]{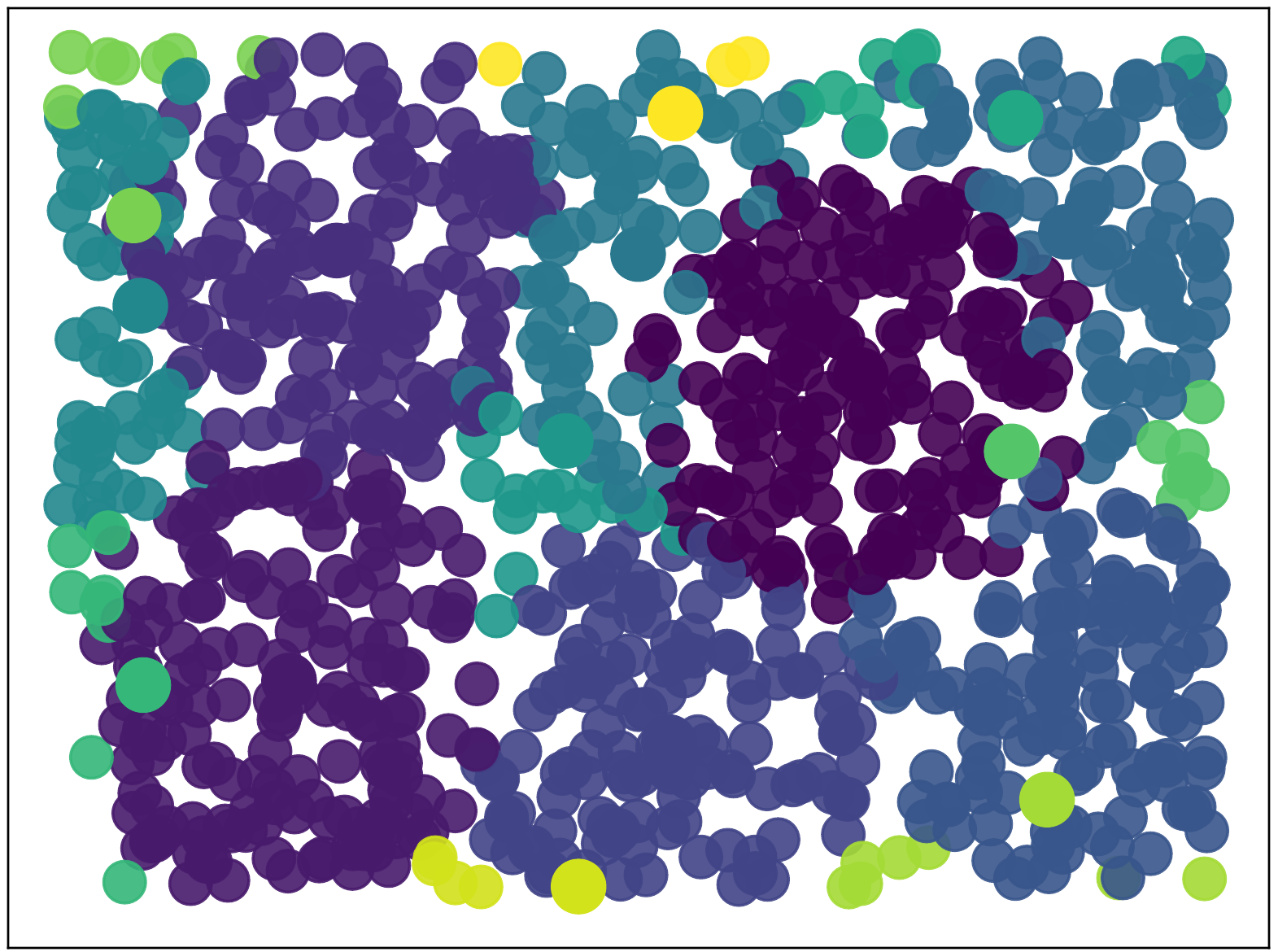}
		\caption{1000 nodes on 5000m $\times$ 7500m area}
	\end{subfigure}
	\begin{subfigure}{.45\textwidth}
		\includegraphics[width=\linewidth]{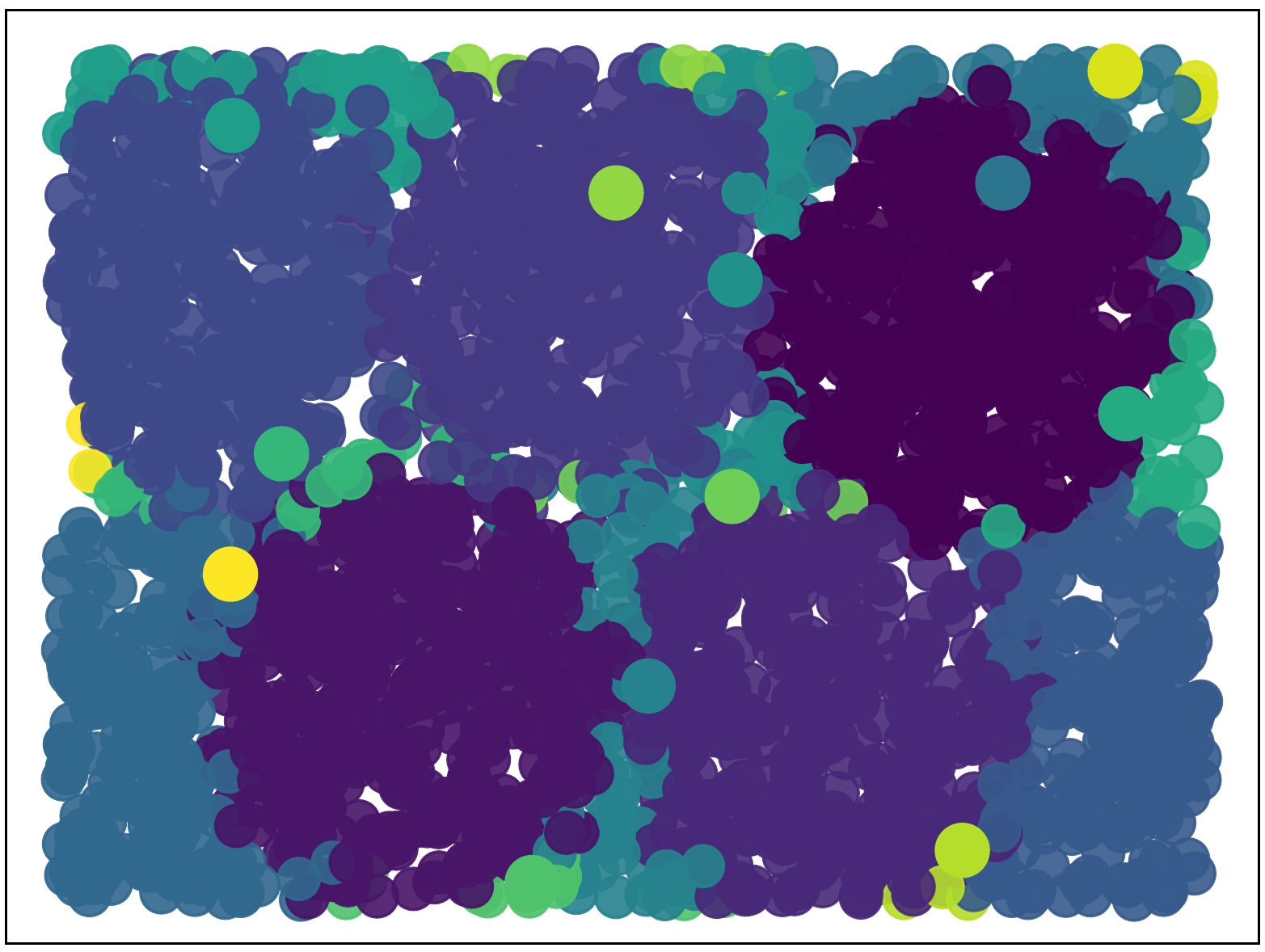}
		\caption{2500 nodes on 5000m $\times$ 7500m area}
	\end{subfigure}
	\caption{Maps of areas served by gateways selected by the proposed algorithm for two sample random topologies for redundancy factor $k=1$.}
\end{figure}

\section{Conclusion}
We have proposed and evaluated an algorithm for sub-optimal selection of LP WAN gateway locations based on $k$-dominating set. The proposed algorithm provides the complete coverage for a given set of nodes, taking into account the maximum number of nodes served by a gateway and required redundancy. The performance evaluation using excesive simulations in both real-life topologies and regular topologies proved that the algorithm is able to select near-optimal locations with a low computing time. 

%

\end{document}